\documentclass[graybox]{svmult}
\usepackage[justification=centering]{caption}
\usepackage{geometry}
\usepackage{mathptmx}       
\usepackage{helvet}         
\usepackage{courier}        
\usepackage{type1cm}        
\usepackage{amssymb}                           
\usepackage{makeidx}         
\usepackage{graphicx}        
\usepackage{multicol}        
\usepackage[bottom]{footmisc}
\usepackage{algorithm}
\usepackage[noend]{algpseudocode}
\usepackage{subfigure}

\makeindex             

\begin{document}

\title*{Ranking and Cooperation in Real-World Complex Networks}
\author{Mohsen Shahriari, Ralf Klamma, Matthias Jarke}
\institute{Mohsen Shahriari \at  Information Systems and Databases
Chair, RWTH Aachen University; \email{shahriari@email.address}
\and Ralf Klamma \at  Information Systems and Databases
Chair, RWTH Aachen University; \email{klamma@dbis.rwth-aachen.de}
\and Matthias Jarke \at  Information Systems and Databases
Chair, RWTH Aachen University; \email{jarke@dbis.rwth-aachen.de}}
%
%
\maketitle


\abstract{People participate and activate in online social networks and thus tremendous amount of network data is generated; data regarding their interactions, interests and activities. Some people search for specific questions through online social platforms such as forums and they may receive a suitable response via experts. To categorize people as experts and to evaluate their willingness to cooperate, one can use ranking and cooperation problems from complex networks. In this paper, we investigate classical ranking algorithms besides the prisoner dilemma game to simulate cooperation and defection of agents. We compute the correlation among the node rank and node cooperativity via three strategies. The first strategy is involved in node level; however, other strategies are calculated regarding neighborhood of nodes. We find out correlations among specific ranking algorithms and cooperativtiy of nodes. Our observations may be applied to estimate the propensity of people (experts) to cooperate in future based on their ranking values.}

\section{Introduction}
\label{Introduction}
Network science has received much attention recently. Advanced computational technologies and machine learning algorithms have created vast opportunities of research in this domain. Moreover, the easy available data sets for network analysis has expanded this area. Ranking algorithms \cite{Klei99} \cite{Page99}, cooperation and defection problem \cite{Nowa06}, link prediction \cite{LiKl04}, overlapping community detection \cite{PDI*05} are lucrative trends in the direction of network analysis. Extracting informative patterns and properties from networks contributes to understand the dynamics behind their formation and evolution.  Moreover, one can support and create online applications out of the found dynamics. One can mention scale-free degree distribution \cite{Watt99b}, small-world-ness \cite{WaSt98d}, densification \cite{LeKF07}, community structures \cite{Newm04} and motifs \cite{MSI*02} as properties of complex networks. 

One key challenge in complex networks is to rank the nodes. In other words, we may require to reliably sort the nodes based on their reputation or other criteria. Ranking algorithms are employed in search engines \cite{Page99}, in addition, they can support some online applications such as expert identification \cite{ShPK15}. Lary Page proposed PageRank that has been used in Google search engine \cite{Page99}. In PageRank, a walker surfs the network based on stochastic random processes. Other ranking algorithms such as HITS and SALSA consider hub and authority vectors to compute rank values of the nodes \cite{Klei99} \cite{LeMo01}. Ranking algorithms are employed to identify experts when people look for such people in question/answer forums. Experts are members of social networks that have higher level of knowledge in certain domains. They help novices to find their intended content online. Classical ranking algorithms and their variations have been employed to identify experts. Their idea is to construct a network among askers and answerers and apply HITS \cite{Klei99} \cite{JuAg07} \cite{JuAg07b} and PageRank \cite{Page99}. Jurczyk and Agichtein constructed a multigraph and connected edges from asker to the question and from question to the answers and from answers to the answerer. However, they omitted the vertices which are questions and answers \cite{JuAg07, JuAg07b}, afterwards, HITS is applied on the resultant graph. Similarly, Zhou et al. assimilate link analysis and topical similarity and extend PageRank to identify experts \cite{ZLLZ12}. They apply Latent Drichlet Allocation (LDA) for topical analysis and extract interesting topics of users. Additionally, a topical PageRank algorithm is employed on the resultant graph. \cite{ZCX*14} \cite{ZCX*11} combine information from various data sources and employ the information from target and relevant categories. They apply random surfer model to rank experts. Although expert identification has been investigated especially with ranking algorithms, task management and cooperativity of experts is still a challenging problem in this regard. Although the expert identification models recommend experts to a certain user, she might not be cooperative and willing to help the user. In this paper, we investigate this challenge via connecting ranking algorithms with cooperation/defection challenge in complex networks.

Cooperation and defection problem has been studied in different domains such as biological, social and animal networks \cite{Nowa06} \cite{MaPr73} \cite{Henr06}. If you pay attention to your daily interaction with people, you figure out that cooperation and defection is a keystone of human societies \cite{OHLN06} \cite{RAB*14} . Even in other networks, we can observe cooperation in different levels. For instance, cells cooperate to form cellular organism and genes cooperate in genomes \cite{Nowa06}. In online social networks cooperation happens in different layers. For instance, if one considers the example of an online learning forum, the cooperation happens between a knowledgeable person (expert) and a person seeking for the information. Additionally, Open Source Software (OSS) development also benefits from volunteer developers in this regard. Nowak mentions five rules for the evolution of cooperation: kin selection, direct reciprocity, indirect reciprocity, network reciprocity and group selection \cite{Nowa06}. In case of OSS, indirect reciprocity through gaining reputation and experience motivates developers to contribute and join OSS projects. In other words, the obtained reputation from one side and the gained experience from the other side inspire developers to join OSS projects 

Cooperation and defection has been majorly studied through Prisoner Dilemma (PD) game in which two players can play an evolutionary game through time. In this game theoretic perspective, each player pays a cost in return for a benefit.  The payoffs are considered as with fitness measure \cite{NoMa92}. There have been some variations to the original PD game such as Tit for Tat, Joss, Friedman and random strategies. In Tit for Tat strategy, an agent cooperates in the first round of the game, afterwards, replicates the behavior of the other player. Joss strategy is based on Tit for Tat; however, it is combined with some random defection in between. Friedman strategy is a bit different and the players usually do not defect at the start of the game but would start not to cooperate as soon as the opponent start defection. Then the player continues defection until the end of the game. In random strategy, each player has a random probability of cooperation and defection which usually is considered equal for the players \cite{KiNa07} \cite{Axel84}.

Cooperation and defection problem has been applied to study correlation among motif frequencies and cooperativity level of motifs \cite{SaRJ10}. A similar problem is to compute the correlation among rank of nodes and their cooperativity level in complex networks. This investigation can be applied in several domains. For instance, it is important to evaluate and predict cooperativity of experts in learning environments or to estimate the probability of joining a developer to a OSS project based on its rank; however, we require a reliable approach to rank the objects in complex networks. To investigate the correlation of cooperativity of nodes and ranking of nodes, we implemented the PD game adopted from Nowak and May \cite{NoMa92} \cite{SaRJ10}. Furthermore, we implemented classical centrality and rank strategies including PageRank, HITS, BetWeen-ness  (BW), CLoseness (CL), Clustering Coefficients (CC) and Simple Degree (SD). Afterwards, three strategies are employed to evaluate the amount of correlation between cooperation and rank problem. First of all, we investigated the correlation of node ranks with cooperativity levels of nodes through Hamming Distance. In other words, each assigned rank value is mapped to a binary value and then compared with the binary cooperativity value obtained through PD game. In the second strategy, we considered the neighborhood of each node and computed the average real rank values of the node neighbors. Moreover, average cooperativity of the node neighborhood is also calculated. Afterwards, these two vectors for the nodes are correlated through relative entropy. Finally, we did the same experiment on the neighborhood of nodes; however, variance of rank values are correlated with variance of cooperativity levels of the neighborhoods.

Through simulations, we could figure out tangible correlations among several ranking algorithms and the cooperation of nodes through PD game. HITS algorithm indicates the highest cooperativity level in comparison to other algorithms in the node level strategy (first strategy). Other algorithms show less correlations and this was observed in almost all the data sets. Regarding the second strategy (mean neighbor rank versus mean neighbor cooperativity), we could observe high correlations regarding the algorithms including SD, CC and CL in most of data sets especially small networks; however, the analysis are more complex and it may require to investigate internal structure of each network. For instance, we only observed tangible correlation when the clustering coefficient of the network is very low. Regarding the third strategy (variance of neighbor rank versus variance of cooperativity rank), we could also identify SD algorithm with the highest correlation among its competitors. In several experiments, we could also observe high correlations for CC and CL  centrality metrics.

The rest of this article is organized as follows. In section \ref{CooperationDefection}, we introduce the basic cooperation and defection game through Prisoner Dilemma and the basic game adopted from Nowak \cite{NoMa92}. In section \ref{RankingApproaches}, classical ranking algorithms are explained in details. In section \ref{SimulationProtocol}, we discuss and mention simulation protocols and employed parameters for the experiments. Next, data sets and evaluation results are explained in section \ref{EvaluationResults}. Finally, we conclude the paper in section \ref{Conclusion}.

\section{Cooperation and Defection}
\label{CooperationDefection}
Network researchers have been interested to understand the dynamics of cooperation and selfish behaviors of individuals and groups. One may not separate interests and benefits of groups from individuals and vice verse. In other words, cooperation of one person with others has reciprocal effects that needs to be studied. In fact, pursuit of selfish agents' behavior might be detrimental for the whole community. These issues have been modeled with game theory approaches such as Prisoner's Dilemma (PD) \cite{RAB*14} \cite{RBH*14}. Merrill M. Flood and Melvin Dresher developed this game in around 1950's that has been under much discussion in computational social and computer sciences. There are some variations to this game; however, the original game which is rampant is a two player game that each agent can take two strategies; either cooperation or non-cooperation. Non-cooperation is also known as defection. The basic mindset behind this problem is that cooperation of both agents leads to higher lucrative outcomes while defection increases individual benefits \cite{SaPa05}.

The original PD game is the story of two persons imprisoned in which the prosecutor can not convict them based on the existing evidence. Here the court specifies a bonus for the pairs and they can behave as follows: both prisoners would serve one year in jail if both stay silent. If one of them betrays, he will be free and the other must serve three years in jail and vice versa. Finally, if both of them betray (defect), they require to serve two years in jail. Rationally speaking, pairwise cooperation serves more beneficial outcomes to both players and nonreciprocal defection is only useful for each of them \cite{EaKl10}.

To simulate the problem, we consider the PD game with a more formal definition. We investigate an evolutionary game played on a network of interactions among individuals or players. Each player can take a set of rational behaviors; named strategies. Here, we mention Cooperation(C) and Defection (D) as possible ones. Based on the selected strategy, she will obtain a payoff; higher payoff values are preferred. Payoff is denoted by a matrix which is shown in the Table \ref{PayoffMatrix}. Elements of the payoff matrix are defined as follows:
\begin{itemize}
\item If both players cooperate then they both receive a payoff $R$.
\item If one cooperates and the other defects then the later receives a $T$ and the former obtains $S$.
\item If both players defect then $P$ is granted to both.
\end{itemize}

Each individual plays with its neighbors in the network and receives the corresponding payoff (fitness). With $T > R >P >S$, the non-cooperation dynamic prevails and reaches a Nash Equilibrium. In contrast, reciprocal cooperation of individuals prevails unilateral cooperation and thus mutual cooperation generates higher payoff \cite{Gint09}.

\begin{table}[h]
\centering
\begin{tabular}{l l l}
\hline
\hline
 & \textbf{C} &  \textbf{D}\\
 \textbf{C} & (R;R) & (S;T) \\
 \textbf{D}  & (T;S) & (P;P) \\
\hline
\end{tabular}
\caption[justification=centering]{Payoff matrix for the basic PD game}
\label{PayoffMatrix}
\end{table}

\section{Ranking Approaches}
\label{RankingApproaches}
To calculate significance and centrality of nodes in a network, we require to apply ranking algorithms. Ranking methods are applied in different application domains such as search engines, recommendation systems and expert identification algorithms. If we denote a graph with $G(V,E)$ in which $V$ is the set of nodes and $E$ is the set of connections between these nodes then a ranking algorithm computes a value assigned to each node. For each node $i$, we denote the set of its neighbors with $d_i$ which stems from the word degree. Moreover, in-neighbors (in-degrees) and out-neighbors (out-degrees) are denoted via $d_i^{in}$ and $d_i^{out}$, respectively. Correspondingly, we employ $\#$ sign to show magnitude of the sets, for instance, $\#d_i$ means number of neighbors of node $i$. In addition, we consider number of nodes as $N$ and number of connections as $M$. Based on the connection context, these rank values can be interpreted as reputation, expertise and social rank; we denote the rank value for node $i$ as $R_i$. Researchers have designed several ranking algorithms that we investigate them in the following.
\subsection{Simple Degree}
Simple degree measure is a classical metric with easy understanding and low computational complexity (O(N)). Simply, we count the number of neighbors of each node ($\#d_i$). Simple degree, which is also called prestige, had good performance in several studies \cite{EaKl10}.
\subsection{PageRank}
In 1990s, Larry Page proposed PageRank which is the most widely discussed ranking algorithm. If we denote PageRank value of each node with $R_i$ then it is updated as follows:
\begin{equation}
R_i= \beta\times \sum\limits_{j \in E_{ji} } {R_j \over {\#d_j^{out}} }+ (1-\beta) \times {1 \over N}.
\end{equation}
Here, $\beta$ is the teleporting parameter showing the probability of starting a new walk by the random surfer \cite{Page99}.
\subsection{Hyperlink-Induced Topic Search}
Hyperlink-Induced Topic Search (HITS) algorithm is proposed by Kleinberg in around 1999 and it was employed to extract useful information from World Wide Web. HITS consists of two vectors named hubs and authorities. Hubs are nodes which refer to other nodes and authorities receive links from other nodes in the network. We initialize the hub and authority vectors with $1\over N$ value and apply the equation  \ref{AuthorityUpdate} and \ref{HubUpdate}, as follows:
\begin{equation}
a_i= \sum\limits_{j \in E_{ji} } h_j, \\
\label{AuthorityUpdate}
\end{equation}
and
\begin{equation}
h_i= \sum \limits_{j \in E_{ij}} a_j,
\label{HubUpdate}
\end{equation}
where $a_i$ and $h_i$ show authority-ness and hub-ness of node $i$. We let the algorithm run until convergence condition is met. Convergent vector $a^*$ is considered as the rank vector ($R$).
\subsection{Closeness Centrality}
Closeness Centrality (CL) shows whether a node is middle of the things \cite{EaKl10}. CL can be defined based on the average shortest distance of a node to all other nodes. It can be written as follows:
\begin{equation}
CL_i= {1 \over \sum \limits_{j=1}^{N} {g^{jk}}}. 
\label{ClosenessCentrality}
\end{equation}
This value can be normalized with $N$ and the $CL$ vector is consider as the $R$ vector.
\subsection{Betweenness Centrality}
BetWeenness Centrality (BW) is a measure which shows how many of nodes should go through you to reach other nodes \cite{EaKl10}. It can be defined as follows:
\begin{equation}
BC_i= {\sum \limits_{j < k} {{g_i^{jk}} \over {g^{jk}}}},
\label{BetweennessCentrality}
\end{equation}
where $g^{jk}$ is the number of shortest paths between node $j$ and $k$ and $g_i^{jk}$ is the number of these paths going through $i$. Similarly, we consider $BW$ vector as the $R$ vector .
\subsection{Clustering Coefficient}
Clustering Coefficient (CC) shows local connectivity and local information exchange in the network. We adopted the definition of CC from Watts and Strogatz \cite{HuGu08} and computed clustering based on the following formula:
\begin{equation}
cc_i= {2(\#E_{d_i})\over \#d_i(\#d_i-1)},
\label{ClusteringCoeffient}
\end{equation}
where $\#E_{d_i}$ is the number of edges between neighbors of node $i$. Here, we consider $CC$ vector equal to $R$ vector.
\section{Simulation Protocol}
\label{SimulationProtocol}
Our goal is to compute the correlation among cooperativity level of nodes and neighbors and their corresponding rank values.
\subsection{Evolutionary Cooperation Process}
In general, cooperative $(\#C)$ level of an evolutionary game on a network of agents interacting through PD game is the final number of cooperative agents. As we want to calculate the correlation between node cooperativity and its rank, we require to know cooperation status of each agent. In this regard, we assign a binary variable as the cooperation status of an agent; 1 denotes cooperation and 0 shows defection. We start the game with equal number of cooperators and defectors. Afterwards in each evolutionary step of the PD game, each agent plays with its neighbors and its payoff value is updated based on the payoff matrix introduced in the Table \ref{PayoffMatrix}. We adopt the parameters for the payoff matrix proposed by Nowak and May and thus we consider $T=b>1, P=S=0, R=1 $ in which $T$ is the propensity to defect \cite{NoMa92}. Synchronously, players update their payoffs; however, through the evolutionary steps of the game, players change their strategies. In other words, each player randomly looks at one of its neighbors and change its strategy with a probability given in the equation \ref{ProbFormula} only if its payoff is less than of its neighbor. In fact, regarding node $i$, it compares its payOff with one random neighbor $j$ and changes its strategy with the following probability:
\begin{equation}
\label{ProbFormula}
		P_{i\Rightarrow j} = { PO_j - PO_i \over {b * maximum(\#d_i,\#d_j)} }.
\end{equation}
To simulate the evolutionary process, we run the PD game for a $TimeWindow=500$. To obtain more stable results, we iterate the PD game 10 times and average over their output. The pseudo code better shows the process. 
\begin{algorithm}
\caption{Evolutionary Prisoner Dilemma Game on Complex Networks}\label{alg:PD-Game}
\begin{algorithmic}[1]
\State $\textit{T} \gets \textit{b}$
\State $\textit{P, S} \gets \textit{0}$
\State $\textit{R} \gets \textit{1}$
\State $\textit{\#Cooperators} \gets N/2$
\State $\textit{\#Defectors} \gets N/2$
\While{\textit{Repetition of the Evolutionary PD Game} $\leqslant$ \textit{Threshold}}
\While{\textit{Iterations} $\leqslant$ \textit{Time Window}}
\For{each node i in the set of all nodes} 
\For{ each node j as neighbor of node i} 
\If{$\textit{strategy(i)} ==C, \textit{strategy(j)==C}$}
\State $\textit{payoffNew(i)} \gets \textit{payoffOld(i)+R}$
\EndIf
\If {$\textit{ strategy(i)==D}, \textit{strategy(j)==C}$}
\State $\textit{payoffNew(i)} \gets \textit{payoffOld(i)+T}$
\EndIf
\EndFor 
\State $\textit{j} \gets \textit{random neighbor of node i}$
\If {$\textit{ payoff(i)}\leqslant \textit{payoff(j)}$}
\State $\textit{strategy(i)} \gets \textit{strategy(j)}$  with $ P_{i\Rightarrow j} = { PO_j - PO_i \over {b * maximum(\#d_i,\#d_j)} }.$
\EndIf
\EndFor

\EndWhile
\State Average over the game realizations
\EndWhile
\Return $\textit{Cooperativitiy Level}$
\end{algorithmic}
\end{algorithm}
\subsection{Node Rank and Cooperativitiy}
To compute the cooperativity level of each node, we consider binary values for each of the ranking algorithms introduced in section \ref{RankingApproaches}. After normalizing each ranking vector $R$ to a real value between 0 and 1, it is mapped to binary values by considering a threshold value equal to mean value of the rank vector $R$. For each of the time windows, we correlate the binary rank vector $R$ with binary cooperation vector $C$. The correlation is simply the Hamming Distance which is the number of binary elements that needs to be changed in order to reach from binary rank vector to cooperation vector .

\subsection{Neighbor Rank and Cooperativitiy}
To observe how cooperativity of a node is related to its rank value, we consider the nodes' neighbors. In other words for node $i$, we compute the average of rank values of neighbors of node $i$ as follows:
\begin{equation}
		R_{i}^{avg}={\sum_{i=1 \in d_i} R_{i}  \over \#d_i},
\end{equation}
where $R_{i}^{avg}$ is the average rank value of node $i$. Similarly variance of rank values of neighbors of node $i$ can be computed as follows:
\begin{equation}
		R_{i}^{var}={{\sum_{i=1 \in d_i} {R_{i}-{R_{i}^{avg}}}^2}  \over \#d_i},
\end{equation}
where $R_{i}^{var}$ is the variance rank values of neighbors of node $i$. Similarly the cooperativity level of neighbors of node $i$ can be calculated as the average of cooperation strategies of nodes $i$. we denote average cooperativity level of neighbors of node $i$ with $\#C_i^{d_i}$. It is calculated as follows:
\begin{equation}
		\#C_i^{d_i}={\sum_{j=1 \in d_i} \#C_j  \over \#d_i}.
\end{equation}
Finally, we correlate $R^{avg}$ and $R^{var}$ vectors with $\#C^{d_i}$ vector based on relative entropy or kullback-leibler divergence (KL divergence). KL divergence is usually applied to compute dissimilarities of two random variables. KL divergence can be computed as follows:
\begin{equation}
		KL(p,q)= { \sum_{k=1}^{K} p_k log({p_k \over q_k}) },
\end{equation}
in which $p$ and $q$ are two vectors that we need to compute their relative information. $KL$ divergence can be rewritten by entropy terms as follows:
\begin{equation}
		KL(p,q)= { \sum_{k=1}^{K} p_k -\sum_{k=1}^{K} p_k log(q_k) }=-H(p)+H(p,q),
\end{equation}
where $H(p,q)$ is named the cross entropy. One can show that when exactly $p=q$ then $KL(p,q)$ equals zero and always $KL(p,q)>=0$ \cite{Murp12}. Here in the evaluations, we compute relative entropy of average and variance neighbor rank and cooperation.
\subsection{Evaluation Results and Discussion}
\label{EvaluationResults}
To observe the correlation among rank values of nodes and their cooperativity level, we employ a couple of real-world data sets as listed in the Table \ref{Datasets}. All of the data sets can be found on the Web.  Some information regarding number of nodes, number of edges, average degree, standard deviation of node degrees and  clustering coefficients of these networks are provided in the corresponding table.

\begin{table}
\centering
 \begin{tabular}{|c|c|c|c|c|c|c|}
  	\hline
  	\textbf{Data set} & \textbf{\# nodes} & \textbf{\# edges} & \textbf{avg} & \textbf{std} & \textbf{CC}& \textbf{Type}\\ \hline
  	Zachary & 34 & 156& 4.59 &15.04& 0.31 &Social\\ \hline
  	 Swamill & 36 & 124& 3.44&4.77& 0.31 &Social \\ \hline
  	 Swamill Strike & 24&76 & 3.17& 1.88& 0.4421&  Social \\ \hline
  	 Dolphin & 62 & 318&5.13 &8.74& 0.26 & Animal \\ \hline
      Gnutella & 6301 & 41554&6.6 & 72.94&0.01&Technological \\ \hline
  	 Email & 1133 & 10903& 9.62 & 87.28 & 0.22 & Communication \\ \hline
      Netscience &1589 & 5485 &3.45&12.04& 0.64& Coauthorship \\ \hline
      Protein & 1706 & 6346 &3.74&48.83&0.0059& Biological \\ \hline
      Euro Roads & 1174 & 2834 &2.41&1.41&0.02& Infrastructure \\ \hline
  	\end{tabular}  	
 \caption{Information regarding the data sets that are used for evaluation.} 	
 \label{Datasets}
\end{table}    

\begin{figure}[ht]
\centering
  \subfigure[Dolphin Network. \label{fig:t0}]{\includegraphics[scale=0.3]{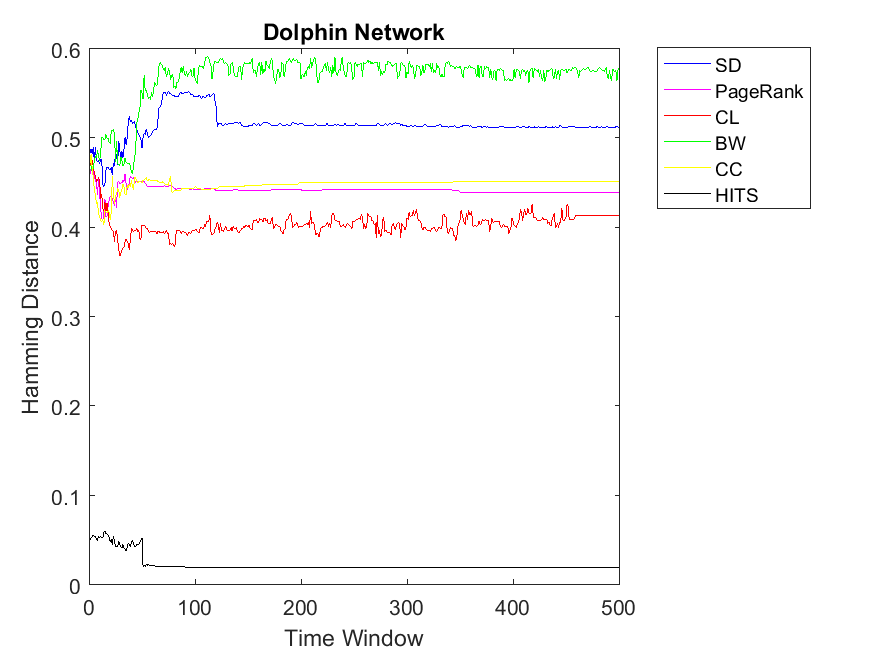}} 
  \subfigure[Zachary Karate Club. \label{fig:t10}]{\includegraphics[scale=0.3]{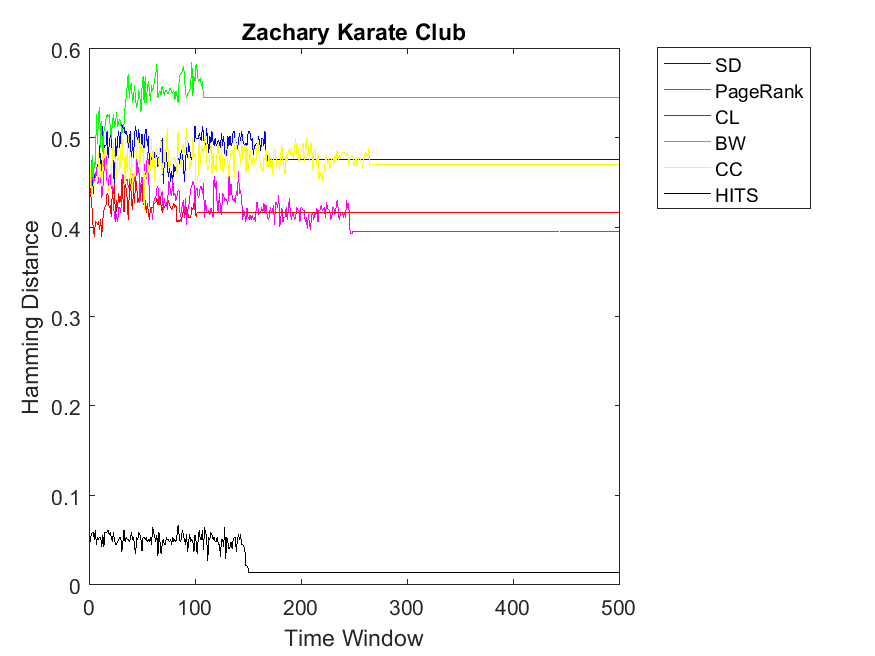}}
    \subfigure[Swamill Network. \label{fig:t0}]{\includegraphics[scale=0.3]{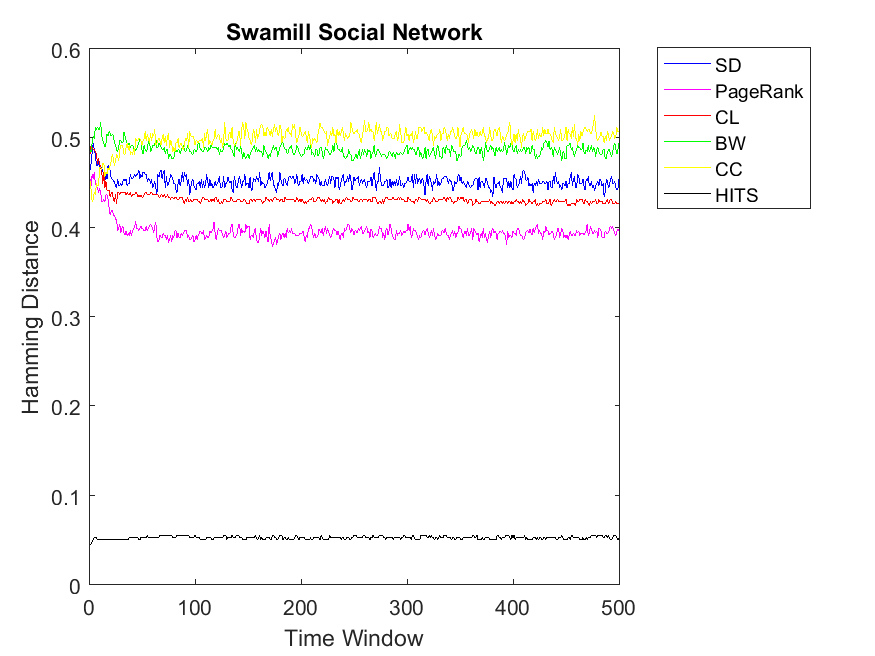}}
    
  \subfigure[Network Science \label{fig:t10}]{\includegraphics[scale=0.3]{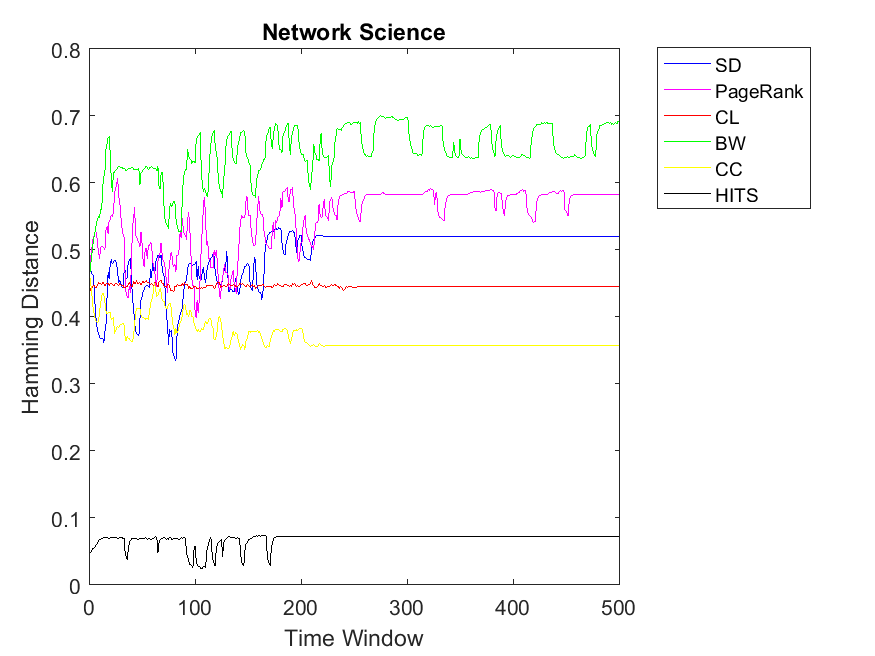}}
\subfigure[Road Network of Europe \label{fig:t10}]{\includegraphics[scale=0.3]{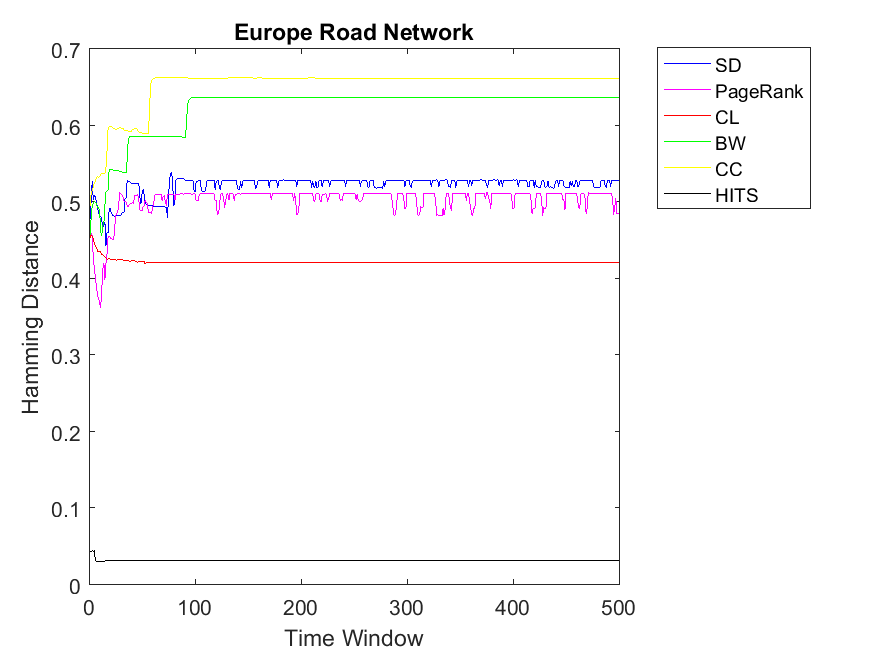}}
\subfigure[Gnutella Peer-to-Peer Network \label{fig:t10}]{\includegraphics[scale=0.3]{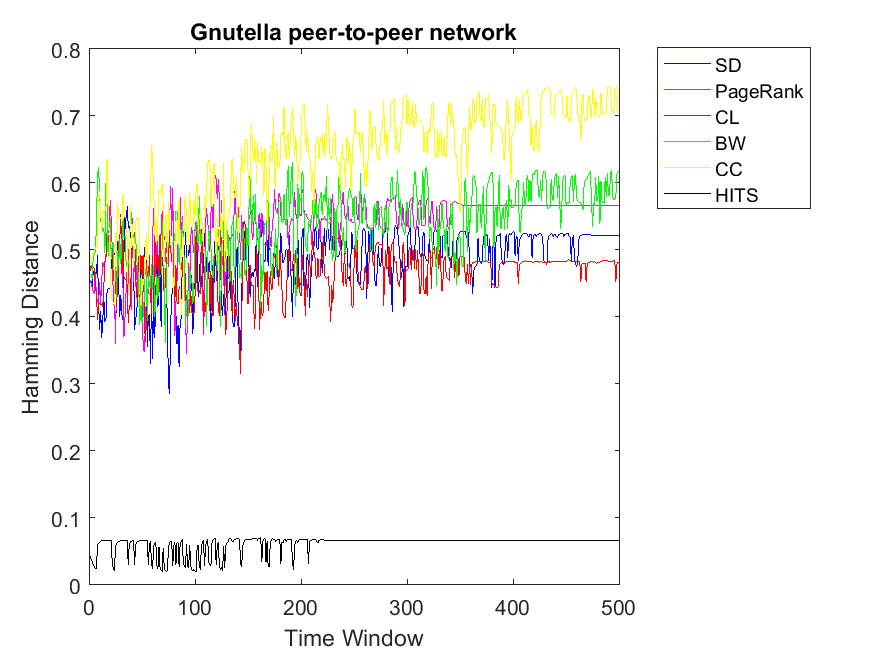}}  
\caption{This figure shows the Hamming Distance of binary node ranks and their cooperativity levels for Dolphin, Zachary, Swamill, Science, Euro road and Gnutella peer-to-peer networks.}
\label{NodeStrategy}
\end{figure}

We apply the three strategies on a couple of data sets and plot the correlation versus the time window of the PD game. For the first strategy, we plot the Hamming Distance of the binary rank vectors and the binary cooperation vector versus number of iterations in the PD game which is set to 500.  Ranking algorithms include Simple Degree (SD), PageRank, Closeness Centrality (CL), BetWeenness centrality (BW), Clustering Coefficient (CC) and HITS. One can observe legends of each algorithm besides its figure.

\begin{figure}[ht]
\centering
  \subfigure[Protein-Protein Interaction Network. \label{fig:t0}]{\includegraphics[scale=0.3]{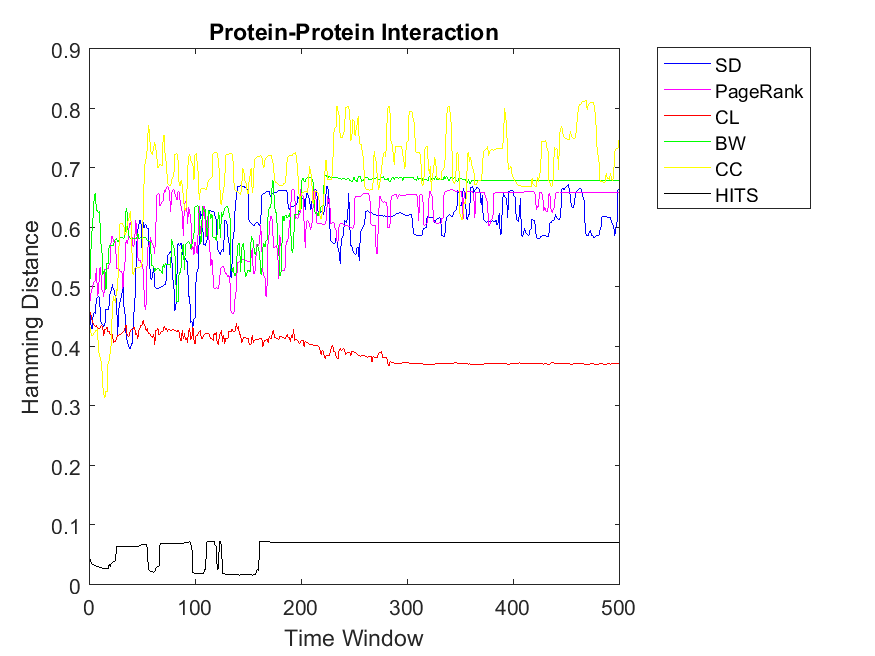}} 
  \subfigure[Email Network. \label{fig:t10}]{\includegraphics[scale=0.3]{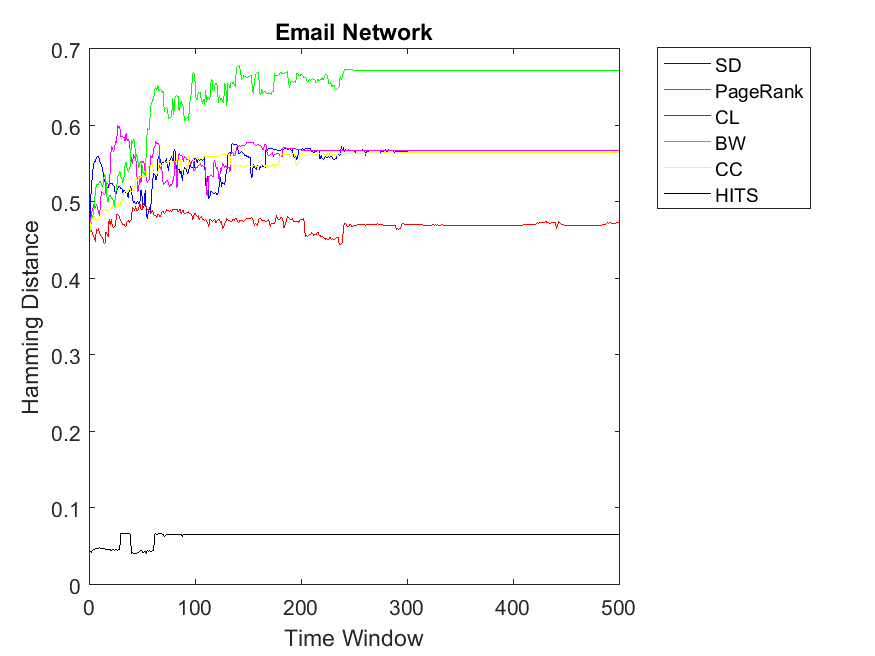}}
  \caption{This figure shows the Hamming Distance of binary node ranks and their cooperativity levels for Protein-Protein Interaction and Email networks.}
  \label{NodeStrategy2}
\end{figure}

Figures \ref{NodeStrategy} and \ref{NodeStrategy2} indicate the correlation among  node ranks and cooperitivity level of each node using Hamming Distance. As it can be observed, the distance is normalized between 0 and 1. Here the higher value indicates less correlation and lower values indicate more similar vectors.  As for Dolphin network, for almost for all the ranking algorithms the correlation values have a smooth horizontal line. This indicates that after a few iterations the PD game on the network convergences to a steady state with a fixed cooperativity level. Moreover, we can observe the highest correlation for HITS algorithm regarding cooperativity. In other words, rank values obtained by HITS have the highest similarity with cooperativity of nodes; mostly less than 0.1 value. In comparison to HITS, other ranking strategies including CL, PageRank, CC, SD and BW obtain less similarity, respectively; however, their difference is not high. In this regard, BW, SD and CC obtain similar distance values of rank and cooperativity. Similar to Dolphin network, we observe the same pattern as for Zachary; HITS showing the highest correlation between node rank and cooperativity. Regarding other ranking algorithms, the patterns might be a bit different; SD has the highest Hamming Distance and BW has the lowest. Additionally, CL, CC and PageRank retain similar correlations. As for Swamill social network, we approximately observe a similar pattern in which HITS shows the highest similarity and others show the least. Regarding Swamill, these correlation values remain between 0.4 and 0.55.

Now we observe bigger networks with higher complexities. Although one can observe some fluctuations, the pattern is still kept the same for network Science data in which HITS has the highest correlation of node rank and cooperativitiy. In this data set, CC and CL have higher similarity values in comparison to the rest and BW produces least correlation values. Regarding the Europe road and email networks, the pattern is smooth and almost the same with HITS reaching the highest similarity of rank and cooperation for nodes. Moreover, other ranking strategies obtain less correlations in the range of 0.4 to 0.75. As for Protein-Protein interaction and Gnutella peer-to-peer networks, the correlations show more fluctuations due to variations of some nodes between cooperation and defection strategies. However, the pattern is kept similar to the previous simulations with HITS obtaining the highest correlation between rank and cooperation.
\begin{figure}[ht]
\centering
  \subfigure[Dolphin Neighbor Mean KL Divergence. \label{fig:t0}]{\includegraphics[scale=0.3]{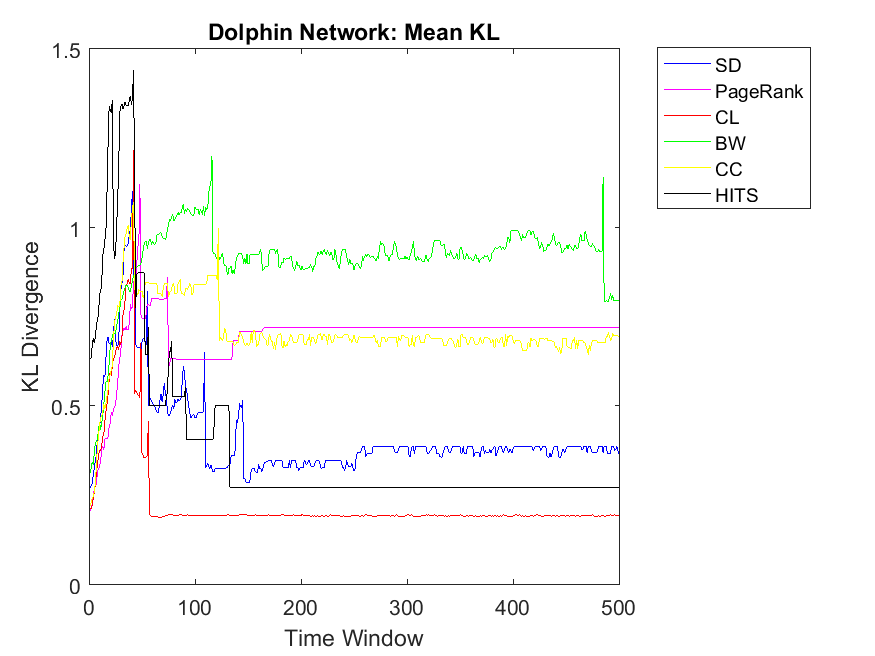}} 
  \subfigure[Dolphin Neighbor Variance KL Divergence. \label{fig:t10}]{\includegraphics[scale=0.3]{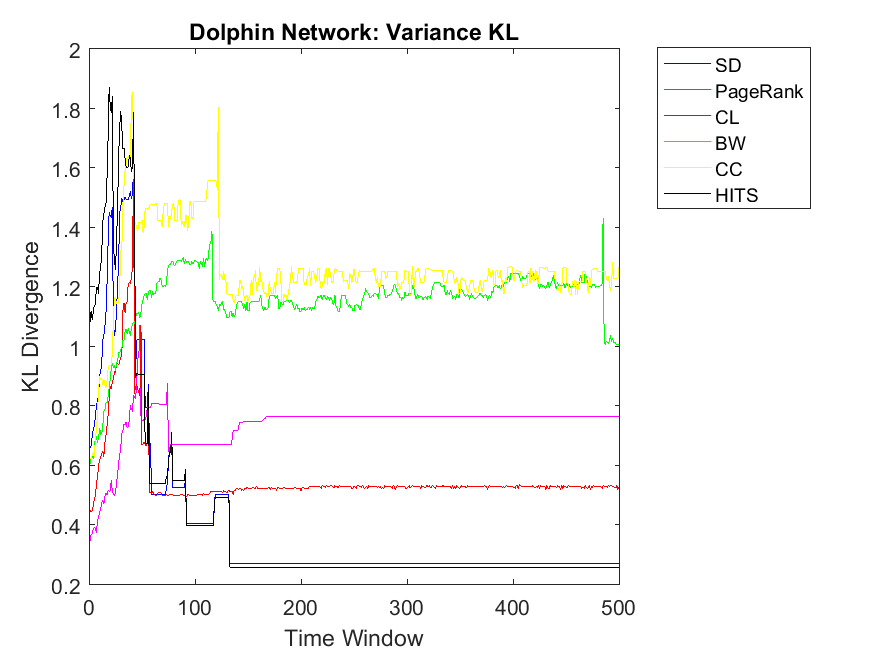}}
   \subfigure[Swamill Neighbor Mean KL Divergence. \label{fig:t0}]{\includegraphics[scale=0.3]{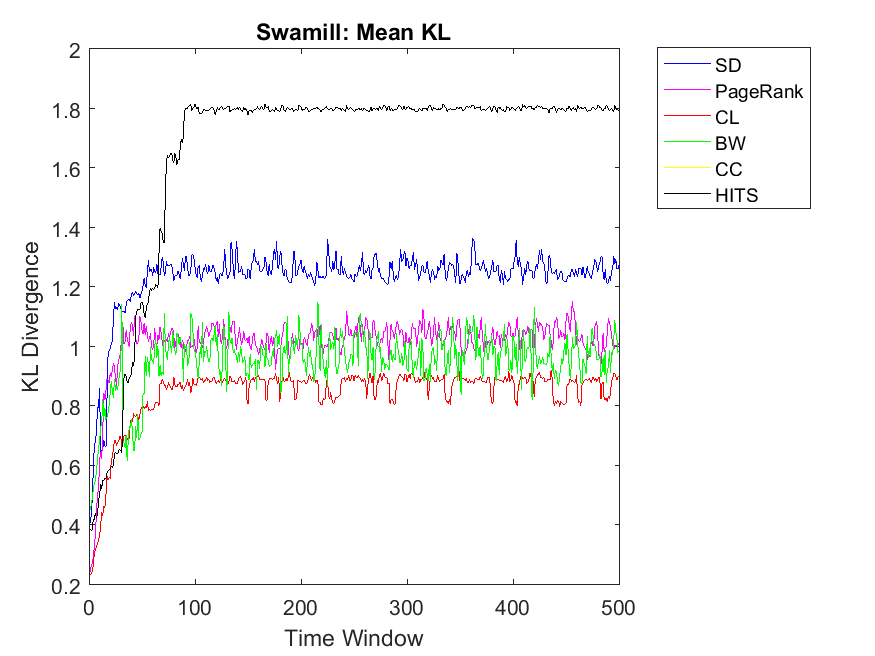}} 
   
  \subfigure[Swamill Neighbor Variance KL Divergence. \label{fig:t10}]{\includegraphics[scale=0.3]{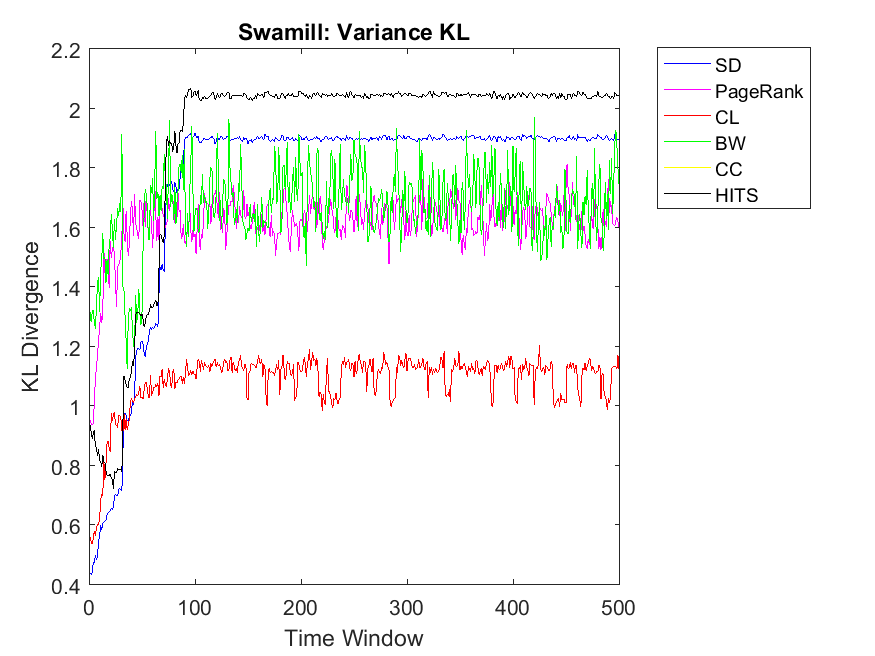}}
   \subfigure[Zachary Neighbor Mean KL Divergence. \label{fig:t0}]{\includegraphics[scale=0.3]{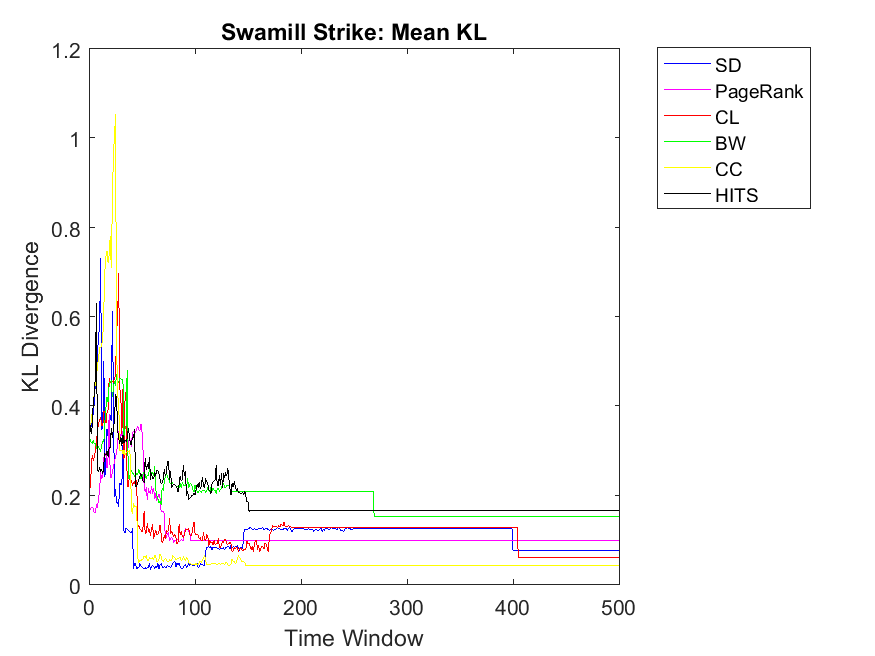}} 
  \subfigure[Zachary Neighbor Variance KL Divergence. \label{fig:t10}]{\includegraphics[scale=0.3]{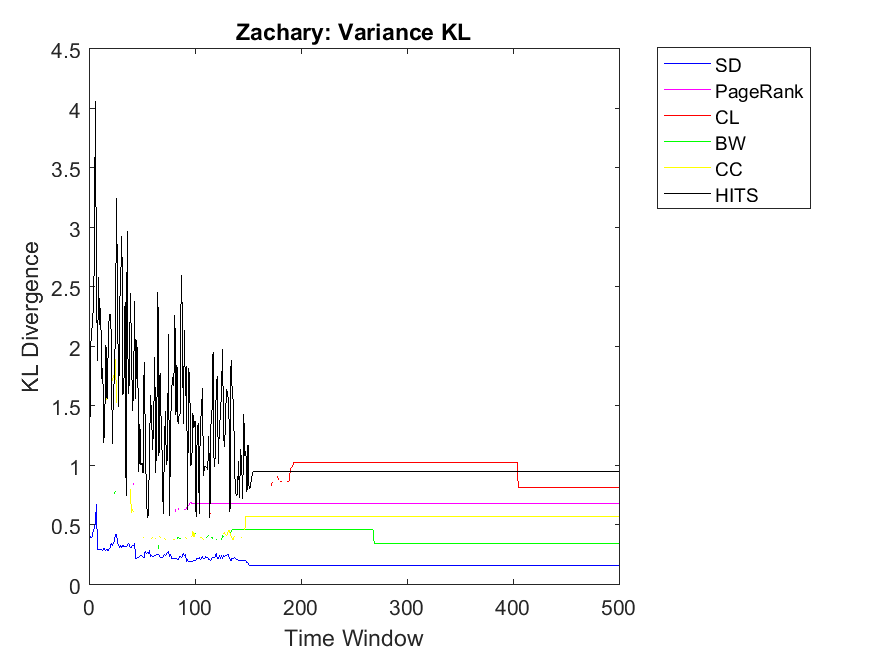}}
  \caption{This figure shows KL divergence of neighbor rank and neighbor cooperativity for mean and variance approaches. It is tested on three data sets including Dolphin, Swamill and Zachary.}
  \label{NeighborStrategy1}
\end{figure}

\begin{figure}[ht]
\centering
  \subfigure[Network Science Mean KL Divergence. \label{fig:t0}]{\includegraphics[scale=0.3]{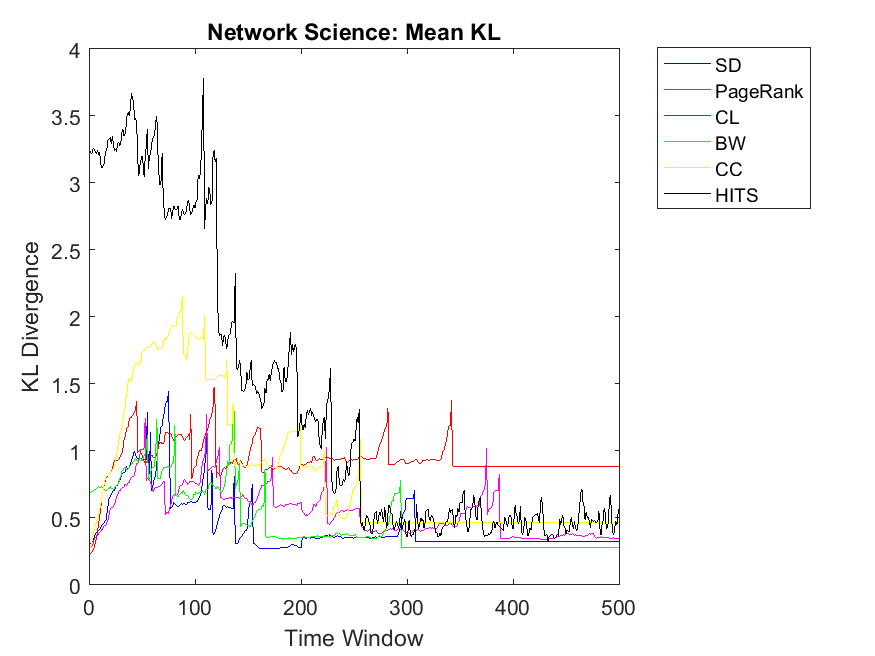}} 
  \subfigure[Network Science Variance KL Divergence. \label{fig:t10}]{\includegraphics[scale=0.3]{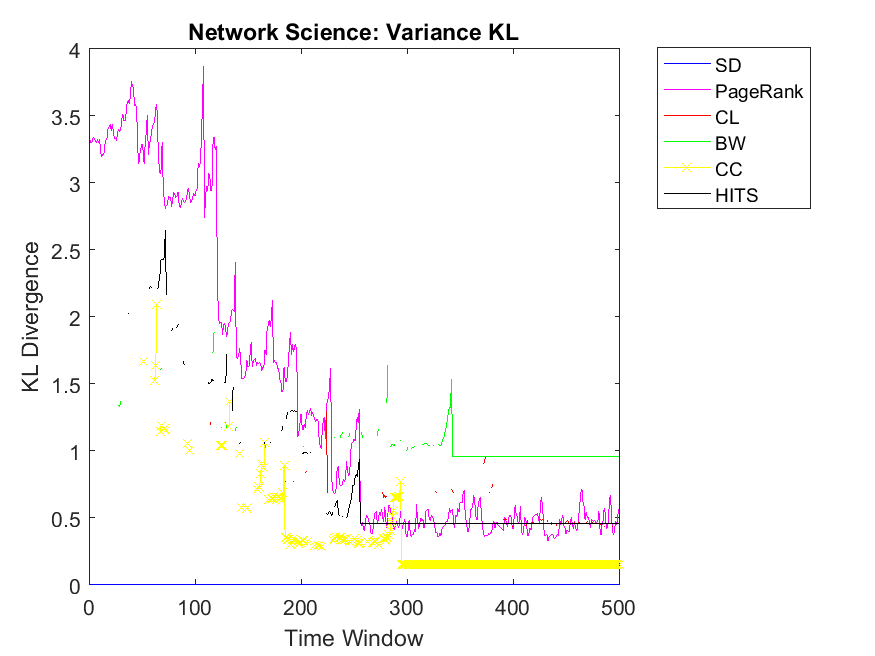}}
  \subfigure[Europe Road Network Mean KL Divergence. \label{fig:t0}]{\includegraphics[scale=0.3]{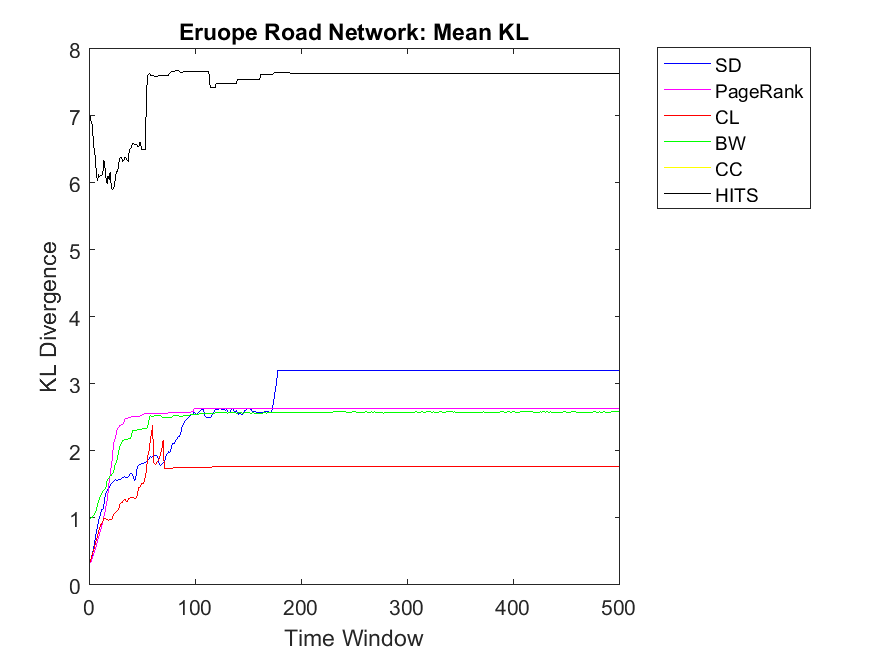}} 
  
  \subfigure[Europe Road Network Variance KL Divergence. \label{fig:t10}]{\includegraphics[scale=0.3]{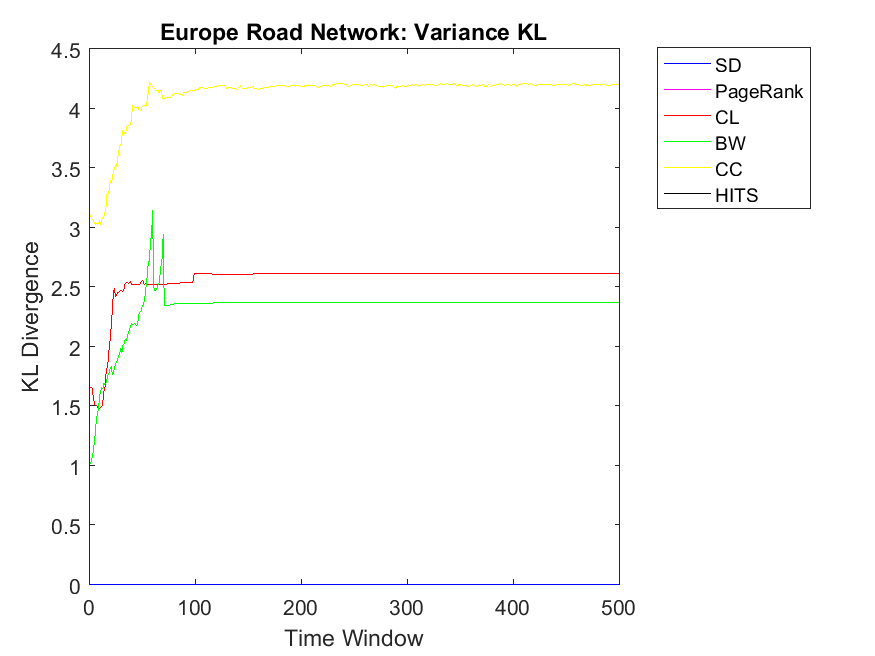}}
   \subfigure[Email Network Mean KL Divergence. \label{fig:t0}]{\includegraphics[scale=0.3]{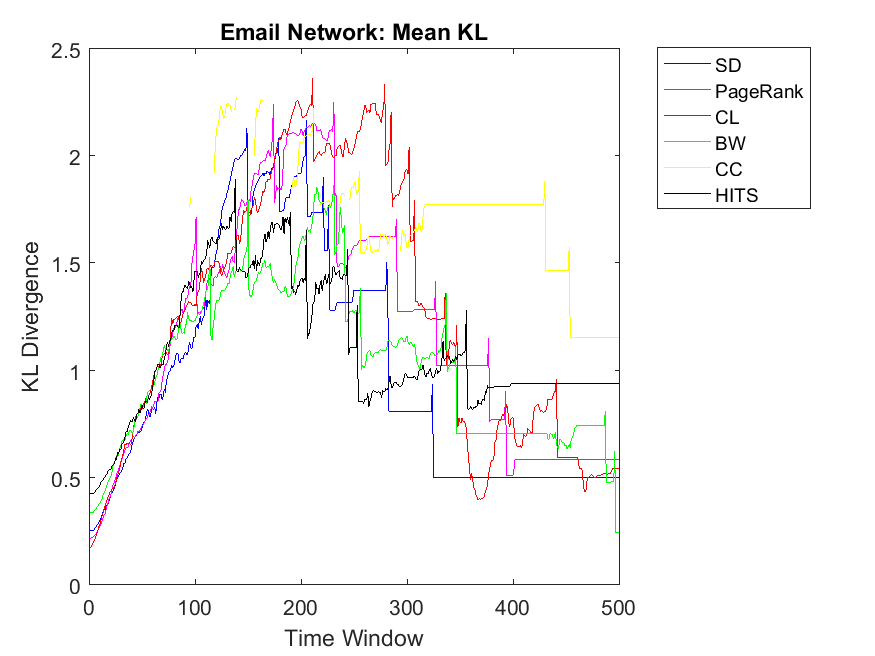}} 
   \subfigure[Email Network Variance KL Divergence. \label{fig:t0}]{\includegraphics[scale=0.3]{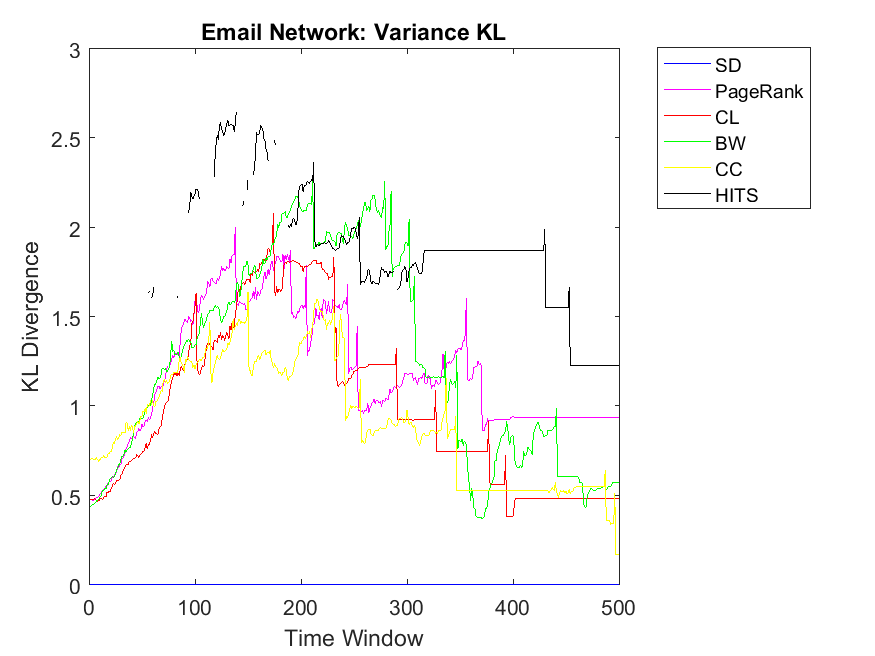}} 
  \caption{This figure shows KL Divergence of neighbor rank and neighbor cooperativity for mean and variance approaches. It is evaluated on three data sets of network science, Europe road and Email Networks. }
  \label{NeighborStrategy2}
\end{figure}

\begin{figure}[ht]
  \centering
  \subfigure[Gnutella Neighbor Mean KL Divergence. \label{fig:t0}]{\includegraphics[scale=0.3]{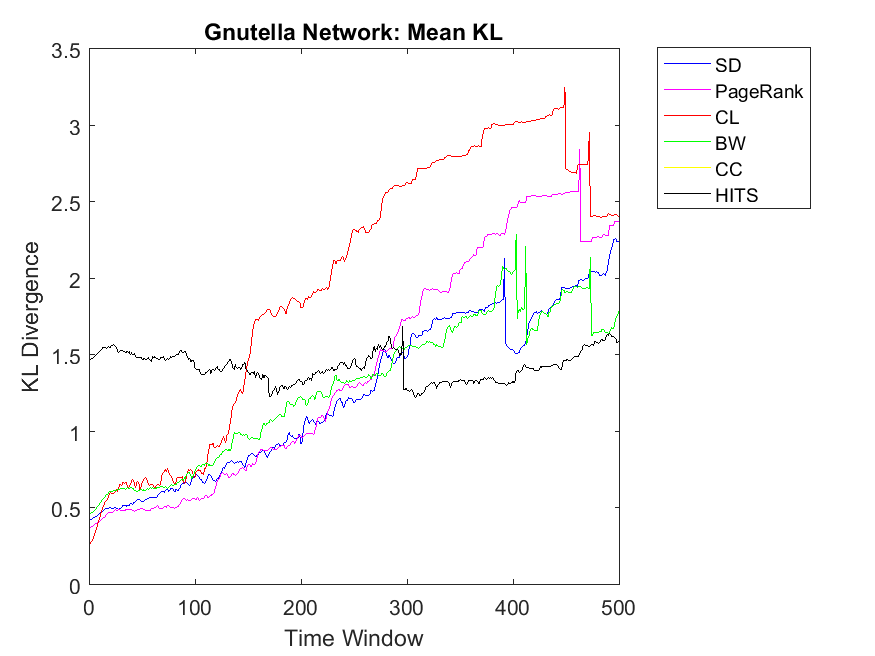}} 
  \subfigure[Gnutella Neighbor Variance KL Divergence. \label{fig:t10}]{\includegraphics[scale=0.3]{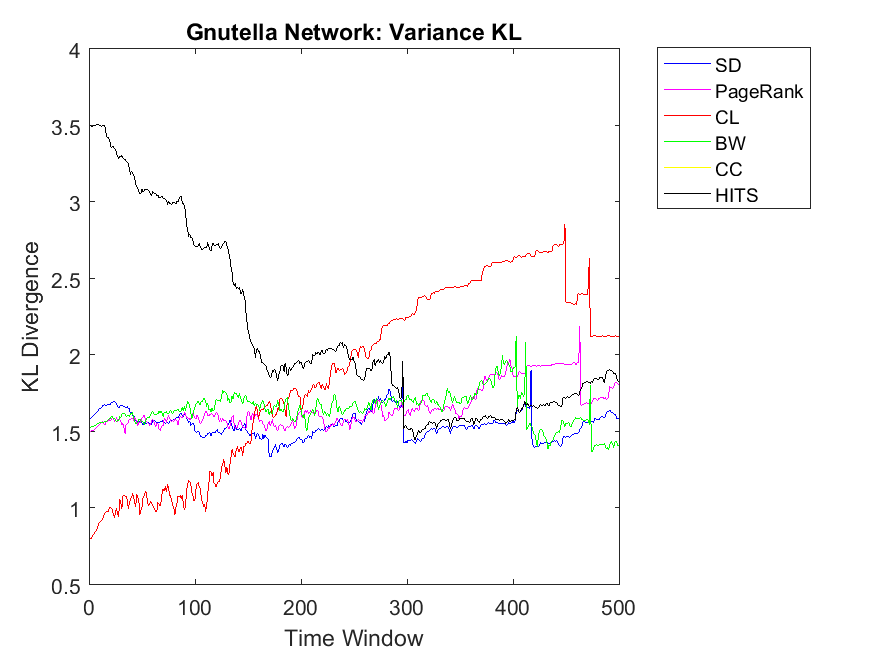}}
    \subfigure[Swamill Strike Neighbor Variance Mean KL. \label{fig:t0}]{\includegraphics[scale=0.3]{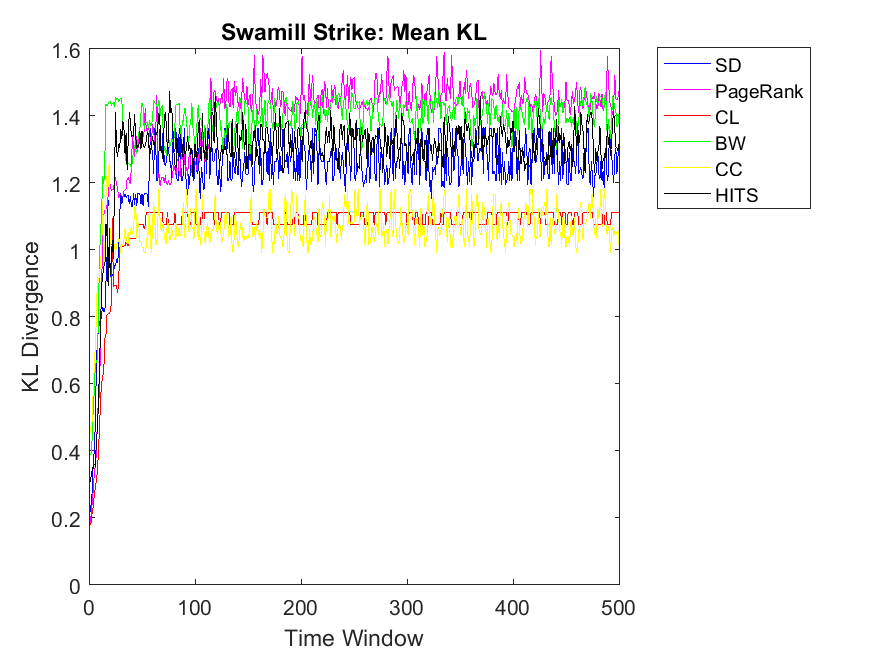}} 
    
  \subfigure[Swamill Strike Neighbor Variance KL. \label{fig:t10}]{\includegraphics[scale=0.3]{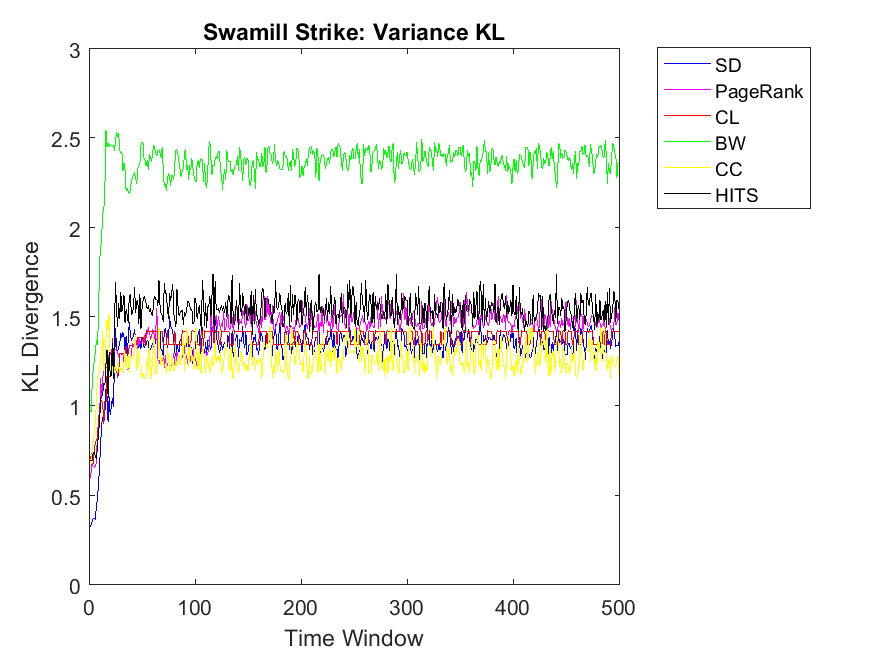}}
     \subfigure[Protein Neighbor Variance Mean KL Divergence. \label{fig:t0}]{\includegraphics[scale=0.3]{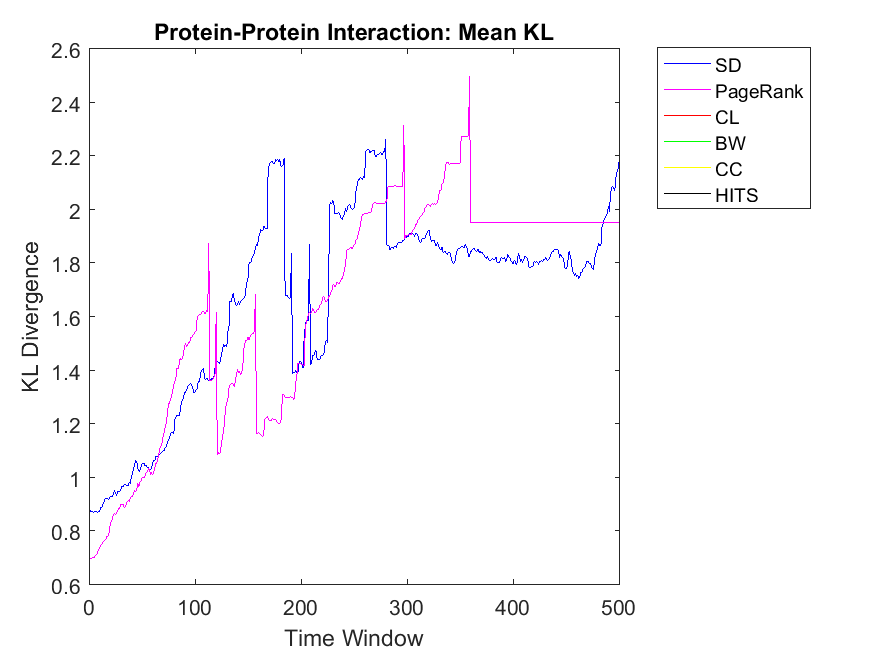}} 
  \subfigure[Protein Neighbor Variance KL Divergence. \label{fig:t10}]{\includegraphics[scale=0.3]{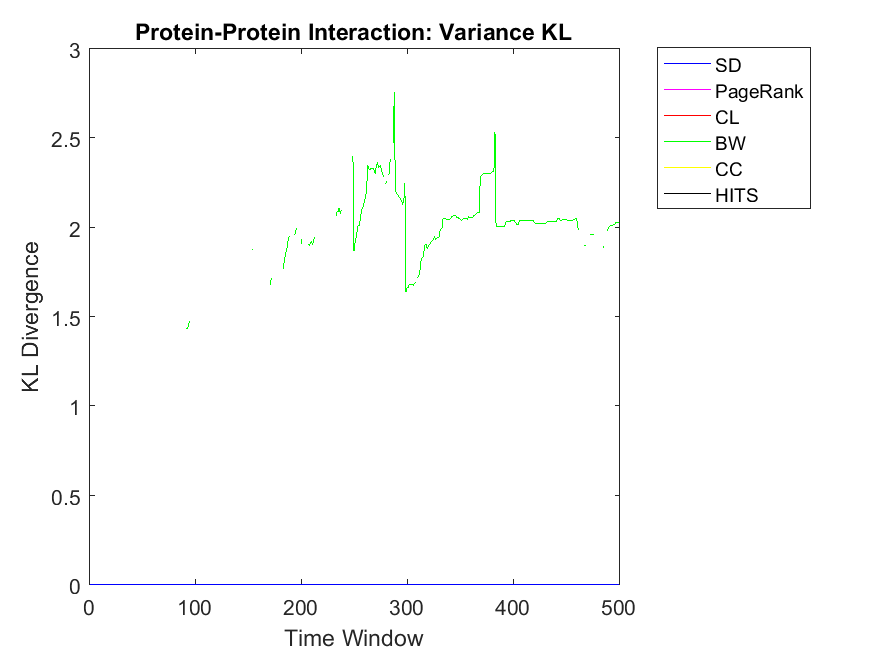}}
  \caption{This figure shows correlation of neighbor rank with neighbor cooperativitiy for mean and variance approaches. It is applied on three data sets of Gnutella, Swamill Strike and Protein-Protein Networks. }
  \label{NeighborStrategy3}
\end{figure}

To extend our observations, we also apply neighbor mean and variance correlation strategies on these data sets. Figures \ref{NeighborStrategy1}, \ref{NeighborStrategy2} and \ref{NeighborStrategy3} show KL divergence of the mean strategy on the left and KL divergence of the variance on the right.  We might notice that KL divergence is a value bigger than zero and can be up to infinite values, therefore, correlations which are very high values are omitted for plotting. When KL divergence approaches zero, it means the neighbor rank and neighbor cooperativity have the highest correlation; this is true for both average and variance values in this regard. Let us first observe smaller data sets, for example Figure \ref{NeighborStrategy1}. Regarding Dolphin network and the mean strategy, the highest correlation can be observed for CL and CC followed by SD, HITS, PageRank and BW, respectively. In other words, if one intends to evaluate the cooperativity level of a node's neighbor with the average rank of the node's neighbor, CL and CC show the highest correlations.  As for Swamill and the Mean strategy, HITS and SD have the highest correlations followed by CL, PageRank, BW, therefore the pattern is a bit different. Regarding Swamill Strike, CL, CC and SD obtain the highest correlation values near to zero which indicates the similarity of mean neighbor rank and mean neighbor cooperation strategies. Here again CC and CL are among highest correlations and also SD is added to previous set. Still looking in to small data sets, Swamill Strike shows the highest correlation for CC followed by CL and SD. Here HITS, BW and PageRank has lower correlations; however, CC and CL were among the top correlation values.

To extend our observation regarding the correlation of mean neighbor rank and mean neighbor cooperation, we investigate bigger data sets to check whether CL, CC and SD show the same correlation patterns. 
Regarding Network science data set, more or less, we can observe that BW and SD obtain the highest similarity followed by CL, HITS, PageRank and CC. As for Europe road network, we can observe almost similar correlations for all the algorithms except HITS with lower similarities. Regarding Email network, the pattern is similar to Europe road network and HITS also shows close similarities like other algorithms. Although HITS showed the highest correlation in node level, it has the lowest correlation for the neighborhood strategy. As for Gnutella network the correlation values are somehow increasing in which it depends on the internal structure of the network. Here BW and HITS have the highest correlation values followed by SD, PageRank, BW and CL. Finally, regarding the Protein network, some correlations were too low that are not plotted in this diagram and the highest correlation can be observed for CL and SD. Overall, we can observe that CL, SD and CC show the highest mean neighbor rank and mean neighbor cooperation correlations; this is also true for big data sets such as Email network and Network science.  By investigating Table \ref{Datasets}, we figure out that Europe road and Science networks and the small data sets such as Swamill, Swamill Strike and Dolphin have approximately low variance degrees; however, data sets such as Gnutella, Email and Protein-Protein interactions have high variance degrees in which their ranking might be different with different strategies. All the data sets have approximately similar average degree for the nodes but the variance degrees can be observed quite differently in which indicates different internal structures.

Regarding neighbor variance strategy, we intend to observe whether CC, CL and SD are among the highest correlation values. As for Dolphin network, SD, HITS and CL have the highest similarities among others. Regarding Swamill, the pattern is a bit different and the highest correlation can be observed for CL followed by PageRank and BW. Here Dolphin has double variance degree in comparison to Swamill which leads into different correlation values. Regarding Zachary, the pattern is smooth and SD, BW and CC obtain the highest correlations among others. Here Zachary has high variance degree in comparison to other data sets. As for Swamill Strike, one  can observe CC, SD, CL, PageRank and HITS among algorithms with high correlation values. Swamill Strike has low variance degree and approximately high clustering coefficient of 0.44 in this regard. As for Europe road network, SD neighbor variance rank vector is completely similar to cooperation neighbor variance vector which is followed by BW. In this regard, the variance degree for Europe road is quite low. As for Email network, we observe the highest variance correlation for SD, CL and CC followed by PageRank, HITS and BW. Correspondingly, Email network has approximately high variance degree of 87.28.
As for Protein, Email, Science and Europe networks, we observe SD as the KL divergence of near to zero which shows that variance simple degree of neighbors of a node has a high correlation with variance of neighbors' cooperativity.  Regarding Protein network, the pattern is a bit different and all other algorithms have no correlation with the cooperativity of the neighbors with the variance strategy. This might be because of the low clustering coefficients observed in the protein networks with 0.0059 value.  Regarding Gnutella with high variance degree, the correlations values are a bit unstable but reaching to a steady state with PageRank, SD and CL as the highest correlations. 

To sum up, HITS showed the highest correlation in the node level strategy; however, centrality metrics such as SD, CL and CC indicated higher correlation among node neighborhood rank and node neighborhood cooperation strategies. Regarding neighborhood strategies, results were more scattered which may require more investigation of internal properties of real-world networks through analyzing overlapping community structures and internal network/community motifs. In addition, we performed the experiments on networks with several thousands of nodes and edges; however, our simulation protocol can be extended on large scale complex networks with millions of nodes and edges. Furthermore, results and the methodology can be applied on different domains such as learning and open source software development environments to estimate the propensity of users (experts) to cooperate through their ranks. Other research domains such as recommender systems and link prediction algorithms may be improved through the joint contribution of rank and cooperativity problems.

\section{Conclusion}
\label{Conclusion}
Cooperation and Defection (CD) is one of the basic phenomenon observed in human interactions. Other network types and biological systems evolve and benefit from this inherent property of complex systems. One can also observe the cooperation in different levels of an online media; in an online question answer forum or an open source software developer community. Nodes in complex networks can be described and categorize by ranking algorithms and our knowledge regarding the cooperativity of a node and its rank value is imperceptible. In this regard, we implemented different ranking algorithms and the basic prisoner dilemma game and applied three strategies to evaluate such correlations. First strategy is related to correlation of the node rank and its cooperativity level. Second and third ones are related to correlation of node neighborhood rank and node neighborhood cooperation. Our results indicate correlation of specific ranking algorithms of each strategy. HITS had high correlation with cooperativity of nodes in the node level strategy. For mean neighborhood strategy the results were more scattered and depends on the network and its internal structure; however, we could mention SD, CC and CL as the ranking metrics with the highest correlations for the second and third strategies. These findings can be applied on expert identification systems for task management and estimating the cooperativity of users through their ranking values. 

\bibliographystyle{acm}

\bibliography{sample.bib}

\end{document}